\documentclass[a4paper,fleqn,12pt]{cas-sc}
\usepackage{soul,xcolor}
\sethlcolor{cyan} 

\usepackage[numbers]{natbib}
\usepackage{amsmath}
\usepackage{longtable}
\usepackage{braket}

\ExplSyntaxOn
\cs_set:Npn \__first_footerline: {}
\ExplSyntaxOff

\begin{document}
\let\WriteBookmarks\relax
\def\floatpagepagefraction{1}
\def\textpagefraction{.001}

\shorttitle{Spectroscopy of Sm$^{3+}$ in K$_2$YF$_5$ }    

\shortauthors{P.\ Chanprakhon et.\ al.}  

\title [mode = title]{Spectroscopy of Sm$^{3+}$ Ions in the C$_{\rm s}$ Symmetry Centres of Hydrothermally Prepared K$_2$YF$_5$ Microcrystals}    

\author[1,2]{Pakwan Chanprakhon}[orcid=0000-0002-0343-8732]
\credit{Investigation, Formal analysis, Writing - original draft}
\author[1,2]{Michael F. Reid}[orcid=0000-0002-2984-9951]
\cormark[1]
\credit{Conceptualization, Software, Investigation, Supervision, Writing - review and editing}
\ead{mike.reid@canterbury.ac.nz}
\author[1,2]{Jon-Paul R. Wells}[orcid=0000-0002-8421-6604]
\cormark[1]
\credit{Investigation, Supervision, Writing - review and editing, Visualization, Resources, Funding Acquisition, Project Administration}
\ead{jon-paul.wells@canterbury.ac.nz}

\affiliation[1]{organization={School of Physical and Chemical Sciences, University of Canterbury},
            addressline={PB4800, Christchurch 8140, New Zealand}}
          
\affiliation[2]{organization={Dodd-Walls Centre for Photonic and Quantum Technologies},
            addressline={New Zealand}}
        
\cortext[1]{Corresponding authors}

\fntext[1]{}


\begin{abstract}
We report on the synthesis and spectroscopic characterization of Sm$^{3+}$-doped K$_2$YF$_5$ microparticles. The particles were synthesized via the hydrothermal technique, yielding a particle size of approximately 20 $\mu$m in length. Scanning electron microscopy (SEM) and X-ray diffraction (XRD) analyses confirmed their orthorhombic crystal structure. A combination of absorption and laser excited fluorescence performed on samples cooled to 10~K, allow for the determination of {56} experimental crystal-field levels. A parametrised crystal-field analysis for Sm$^{3+}$ in the C$_{\rm s}$ point group symmetry centres of K$_2$YF$_5$ yields good approximation to the data.

\end{abstract}

  



\maketitle

\section{Introduction}

Lanthanide doped micro- and nano-particles have gained significant attention for the prospect of diverse applications in biomedical imaging, optical sensing, anti-counterfeiting, lighting, and display technologies \cite{Dong2015, Bouzigues2011, hao2013sensing, zhao2016design}. These materials offer favourable optical and mechanical properties, as well as being largely non-toxic.

Among the lanthanide ions, Sm$^{3+}$
is not always the first choice in optical applications and very rarely so for the development of lasers. The 4f$^5$ configuration offers a rich energy level structure with the $^6$H and $^6$F Coulombic terms providing a plethora of levels from which many pathways are available for energy transfer cross-relaxation of the lowest lying metastable 'optical' level $^4$G$_{5/2}$; at typical dopant concentrations used for solid state lasers. {Consequently}, comparatively high energy optical pumping is required for optimal excitation efficiency at around 405 nm, corresponding to excitation of the $^6$H$_{5/2}\rightarrow ^6$P$_{3/2}$, $^6$P$_{5/2}$ transitions. However, Sm$^{3+}$ exhibits strong $^4$G$_{5/2}\rightarrow$$^6$H$_{7/2}$ transitions, which produce intense orange fluorescence. The first Sm$^{3+}$ laser was demonstrated in 1979 using a flashlamp pumped TbF$_{3}$ host crystal at just above 110 ~K \cite{kazakov1979}. In very recent years, laser action has been demonstrated in LiYF$_{4}$, LiLuF$_{4}$, KGd(WO$_{4}$)$_{2}$ and YAlO$_{3}$ doped with Sm$^{3+}$ using III-V nitride multi-quantum well laser diodes or doubled, optically pumped semiconductor lasers (diode pumped VCSELs) as pump sources in the 460-480 nm region \cite{marzahl2015spectroscopy, Demaimay2025, demaimay2024orange, baillard2023, kaneda2025, chu2025}. Of the hosts for which laser action has been achieved, LiYF$_{4}$ is arguably the most well studied, see for example \cite{wells1999, hui2004investigations, duffy1999bridgman, yamaga2012resonant}; in fact for LiYF$_{4}$ the lifetime of the terminal state of the lasing transition has been measured as 3.1 picoseconds \cite{horvath2017} - nine orders of magnitude faster than the upper state lifetime. However, other host materials are certainly worthy of investigation in this context.
 
Potassium yttrium pentafluoride (K$_2$YF$_5$) has good thermal stability, optical transparency, and relatively low phonon energies \cite{tuyen2020k2yf5, hanh2010thermoluminescence, wang2019synthesis}.
K$_2$YF$_5$ crystallizes in the orthorhombic Pnam space group, where lanthanide ions substitute at a single lattice site with minimal structural disruption. Each Y$^{3+}$ ion is coordinated by seven fluoride ions, yielding a centre having \textit{low} C$_{\rm s}$ point group symmetry \cite{loncke2007k}, in contrast to the tetragonal symmetry centres in Scheelite structure LiYF$_{4}$.  Some of the relevant optical properties of Sm$^{3+}$- doped K$_2$YF$_5$ have been reported previously \cite{khaidukov2021study, van2012judd}, however, a detailed investigation of the electronic structure is, as yet, unreported.

In this work, we report a detailed experimental and computational spectroscopic study of Sm$^{3+}$ doped K$_2$YF$_5$. We have chosen to use microparticles for this purpose because of their ease of preparation. The samples were prepared using the hydrothermal technique, and their crystal structure and morphology were analyzed using XRD and SEM. Low-temperature absorption and laser excitation and fluorescence, coupled with parametrised crystal-field calculations, were employed to determine the 4f$^5$ electronic energy levels.

\section{Materials and Methods}

\subsection{Synthesis of K$_2$YF$_5$:Sm$^{3+}$ Microparticles and Characterization}

The K$_2$YF$_5$ samples were synthesized using a hydrothermal technique, following the procedure detailed in previous work (see reference \cite{Bian2019, chanprakhon2025}). Sm(NO$_3$)$_3$·H$_2$O and Y(NO$_3$)$_3$·6H$_2$O were used as precursors. In this study, samples doped with 2.5\% Sm$^{3+}$ were prepared and used for all measurements; a comparable concentration to that used for laser materials.

\subsection{Characterisation}

X-ray diffraction (XRD) data was collected using a RIGAKU 3 kW SmartLab X-ray diffraction spectrometer, operating at 40 kV and 30 mA. Morphological images were captured using a scanning electron microscope (SEM) JOEL JSM IT-300 LV, and the particle size distribution was analyzed with the ImageJ program.  

\subsection{Absorption and Fluorescence}

Absorption spectra were measured using a N$_{2}$ gas purged, Bruker Vertex 80 FTIR, having a maximum apodised resolution of 0.075 cm$^{-1}$. A thin pellet of K$_2$YF$_5$ doped with 2.5\% Sm$^{3+}$ was mounted on a copper sample holder, which was connected to the cold finger of a Janis closed-cycle helium compressor. The absorption measurements were conducted at a nominal temperature of 7 K, using a combination of MCT, InGaAs, and silicon photodetectors.

For the laser spectroscopy, measurements were conducted at a nominal sample temperature of 10 K with a Janis closed-cycle helium compressor. The samples were excited with a PTI wavelength-tunable pulsed dye laser, utilizing Coumarin 500 and C540A dyes. Emission and excitation spectra were recorded through an iHR550 single monochromator and Peltier-cooled Hamamatsu R2257P and H10330C photomultiplier tubes with photon counting and under computer control.

\subsection{Crystal-Field Calculations}

We have previously reported calculations for K$_2$YF$_5$:Er$^{3+}$ \cite{Solanki2024} and K$_2$YF$_5$:Ho$^{3+}$ \cite{chanprakhon2025}. As for Er$^{3,+}$, Sm$^{3+}$ has an odd number of 4f electrons, so in the absence of magnetic fields all electronic states are doubly degenerate. 
We use a parametrised Hamiltonian \cite{wybourne1965spectroscopic, Carnall1989, reid2016theory}:

\begin{equation}
    H = H_\text{FI} + H_\text{CF}
\end{equation}
where $H_\text{FI}$ is the free ion contribution and $H_\text{CF}$ is the crystal-field contribution. Detailed information on the free-ion Hamiltonian is available in the references cited above.

The crystal-field Hamiltonian can be expressed using Racah spherical tensors as follows.:

\begin{equation}
\label{eq:cf}
    H_\text{CF} = \sum_{k,q} B_{q}^{k}C_{q}^{(k)}, 
\end{equation}
where the $B_{q}^{k}$ and the $C_{q}^{(k)}$ are crystal-field parameters and Racah spherical tensor operators for the 4f$^N$ configuration, respectively. The crystal-field parameters, $B^k_q$, \( k \) are restricted to  \( k \) = 2, 4, 6. Due to the C$_s$ symmetry of the Y$^{3+}$ sites in K$_2$YF$_5$, the $B^k_q$ are zero when $q$ is odd and are imaginary when $q>0$. 

\clearpage
\section{Results and Discussion}
\subsection{Phase, Morphology, and Composition}

\begin{figure}[tb!]
\centering
 \includegraphics[width=0.5\textwidth]{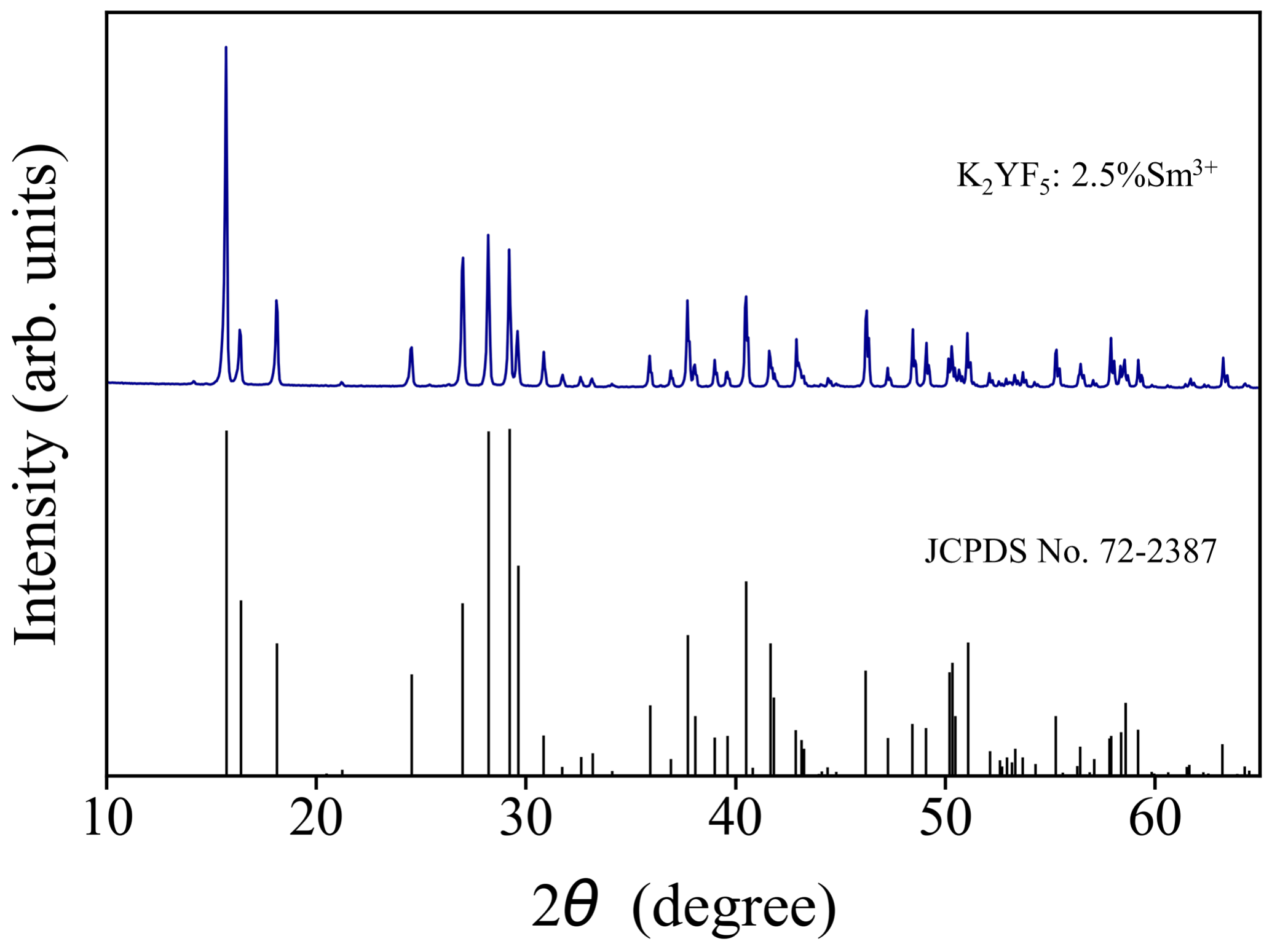} 
\caption{ \label{fig:xrd}
XRD pattern of K$_2$YF$_2$ microparticles with a Sm$^{3+}$ ion dopant concentration of 2.5 molar percent}
\end{figure}

\begin{figure}[tb!]
\centering
 \includegraphics[width=1\textwidth]{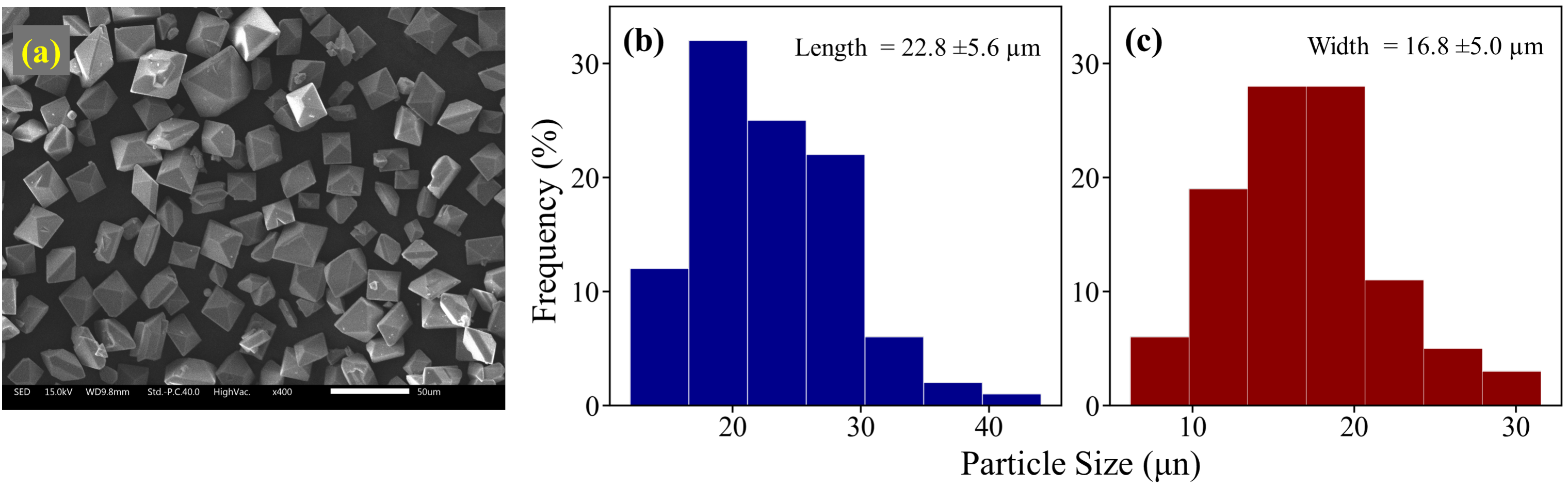} 
\caption{ \label{fig:SEM}
(a) SEM image of K$_2$YF$_5$ doped with 2.5\%Sm$^{3+}$ (b),(c) The size distribution of the microparticles measured from length and width, respectively.}
\end{figure}

The powder XRD patterns of the K$_2$YF$_5$ sample are presented in Figure \ref{fig:xrd}, alongside reference data from the JCPDS No.77-2387 database. The XRD pattern for the synthesized sample exhibits good agreement with the reference data, confirming the successful synthesis of the desired phase.
Figure \ref{fig:SEM} illustrates the morphology of 2.5\% Sm$^{3+}$-doped microparticles, revealing an octahedral shape. The particle size distribution is quite uniform, with an average length of 22.8 $\pm$ 5.6 $\mu$m and a width of 16.8 $\pm$ 5.0 $\mu$m. 

\clearpage

\subsection{Absorption and Laser-Excited Fluorescence Measurements}
 
\begin{figure}[tb!]
\centering
 \includegraphics[width=0.8\textwidth]{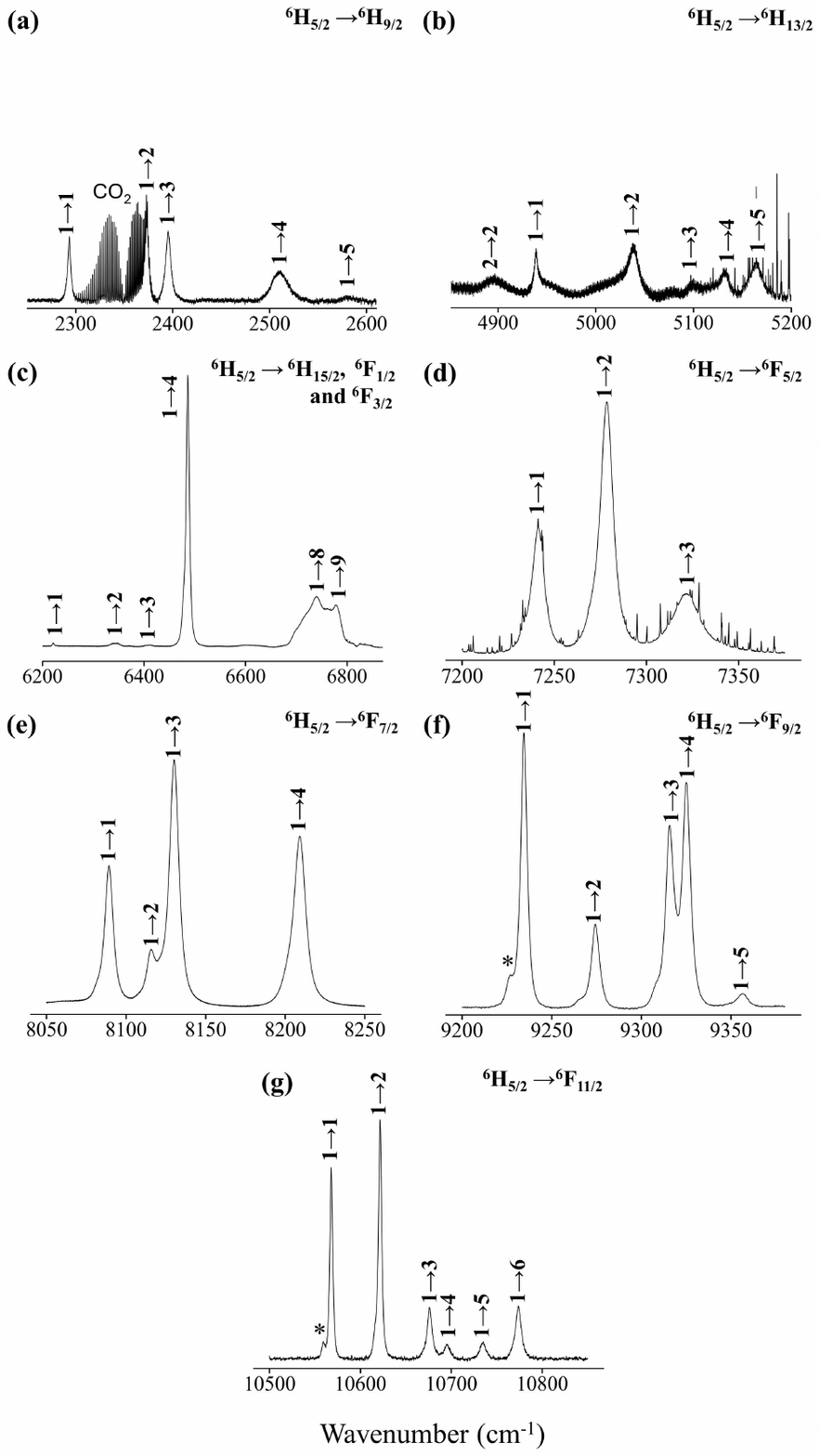} 
\caption{ \label{fig:absorption}
The absorption spectra of K$_2$YF$_5$: 2.5\% Sm$^{3+}$ microparticles measured at 7 K, showing the transitions for
(a) $^6$H$_{5/2}$ $\rightarrow$  $^6$H$_{9/2}$;
(a) $^6$H$_{5/2}$ $\rightarrow$  $^6$H$_{13/2}$;
(c) $^6$H$_{5/2}$ $\rightarrow$  $^6$H$_{15/2}$, $^6$F$_{1/2}$ and $^6$F$_{3/2}$;
(d) $^6$H$_{5/2}$ $\rightarrow$  $^6$F$_{5/2}$;
(e) $^6$H$_{5/2}$ $\rightarrow$  $^6$F$_{7/2}$;
(f) $^6$H$_{5/2}$ $\rightarrow$  $^6$F$_{9/2}$;
(g) $^6$H$_{5/2}$ $\rightarrow$  $^6$F$_{11/2}$.
* indicates an unassigned spectral feature.}
\end{figure}

The absorption spectra of a K$_2$YF$_5$ pellet doped with 2.5\% Sm$^{3+}$ were measured at a nominal sample temperature of 7 K. Figure \ref{fig:absorption} shows the absorption spectra of 9 multiplets in the mid- to near-infrared arising from transitions from the $^{6}$H$_{5/2}$ ground state to nine higher lying multiplets of the $^{6}$H and $^{6}$F terms. Sharp spectral lines due to both atmospheric water and CO$_{2}$ are present in the spectra due to incomplete purging of the entire optical beam path.

\begin{figure}[tb!]
\centering
 \includegraphics[width=0.9\textwidth]{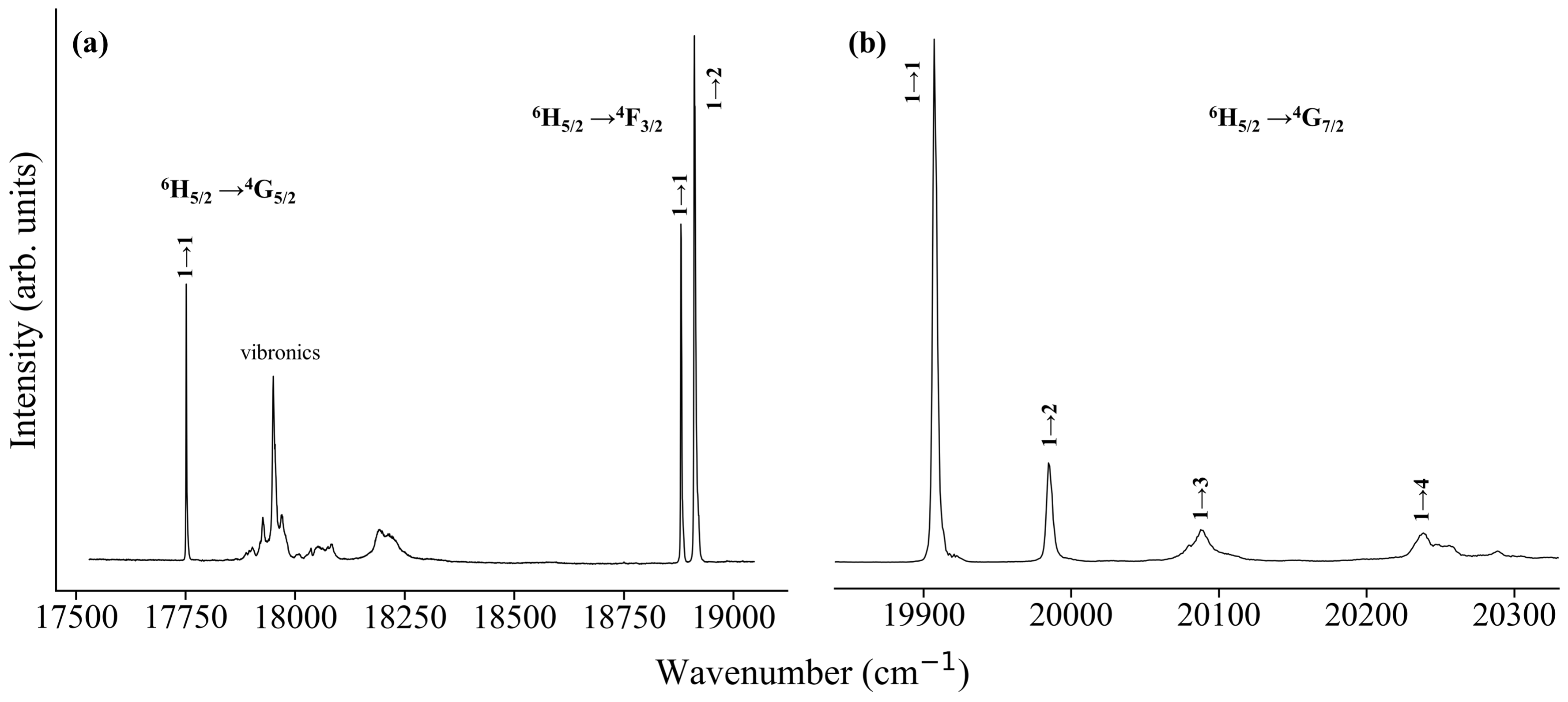} 
\caption{ \label{fig:excitation}
 The excitation spectrum of K$_2$YF$_5$: 2.5\% Sm$^{3+}$ microparticles measured at 7 K, showing (a) the $^6$H$_{5/2}$ $\rightarrow$  $^4$G$_{5/2}$ and $^4$F$_{3/2}$ transitions, and (b) the $^6$H$_{5/2}$ $\rightarrow$  $^4$G$_{7/2}$.}
\end{figure}

Figure \ref{fig:excitation} presents the laser excitation spectra for the $^{4}$G$_{5/2}$, $^{4}$F$_{3/2}$, and $^{4}$G$_{7/2}$ multiplets, obtained using Coumarin 540A for $^{4}$G$_{5/2}$ and $^{4}$F$_{3/2}$ multiplets, and Coumarin 500 for $^{4}$G$_{7/2}$ multiplet as the gain medium in the pulsed dye laser and with the sample cooled to 7~K. Fluorescence was monitored at 16,652 cm$^{-1}$, corresponding to the magnetic dipole allowed  $^{4}$G$_{5/2}$A$_{1}\rightarrow ^{6}$H$_{7/2}$Y$_{1}$ transition.
Vibrational sidebands have been observed for the $^{4}$G$_{5/2}$ multiplet in several hosts \cite{wells1999, wells2000, wells19991, horvath20191}
and we observe strong vibrational sidebands that broadly agree with the Raman data from \cite{tuyen2020k2yf5}.
The presence of these sidebands make the assignment of electronic states uncertain, apart from the lowest state.

\begin{figure}[tb!]
\centering
 \includegraphics[width=1\textwidth]{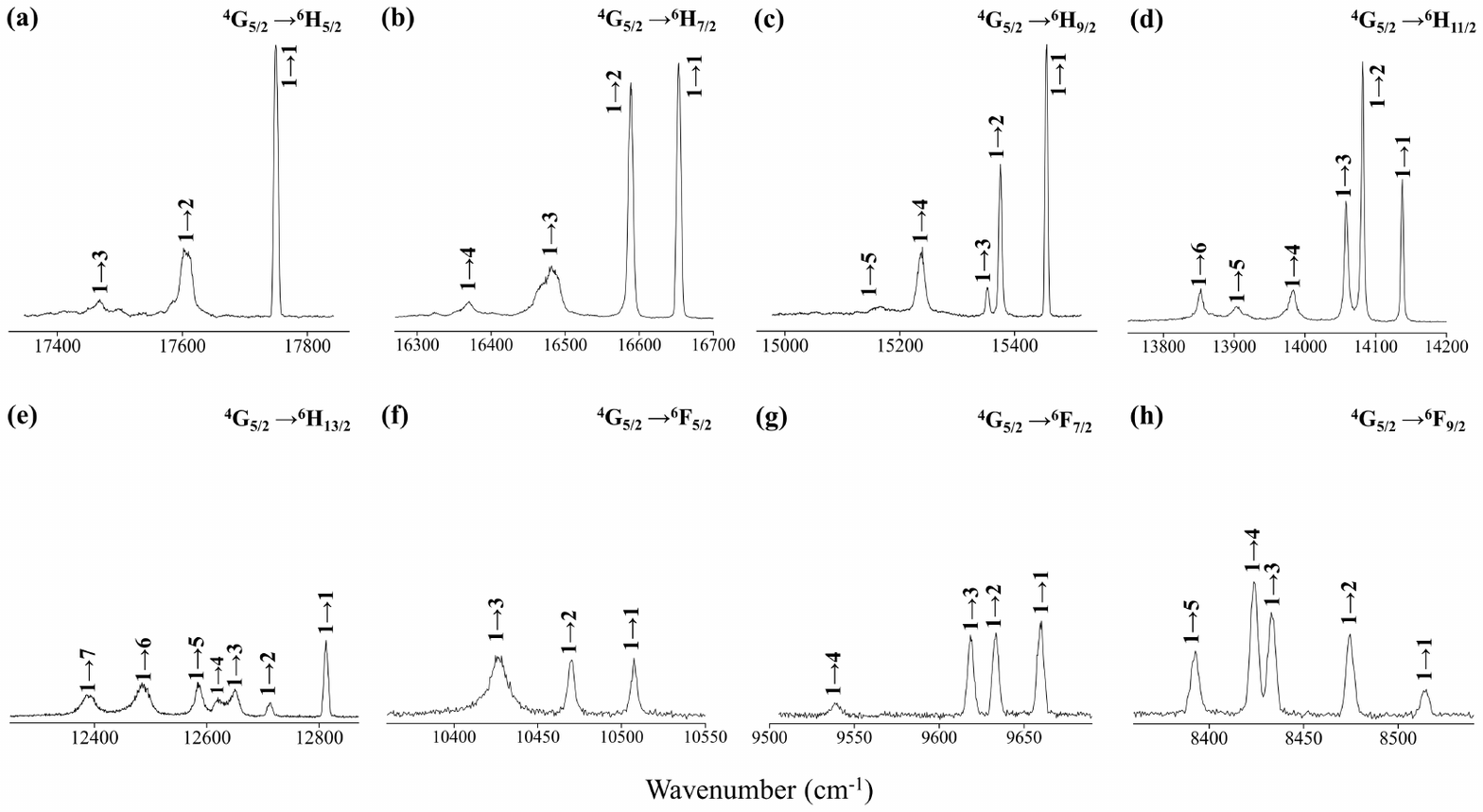} 
\caption{ \label{fig:emission}
The fluorescence spectra of K$_2$YF$_5$: 2.5\% Sm$^{3+}$ microparticles measured at 7 K, showing the
(a) $^4$G$_{5/2}$ $\rightarrow$  $^6$H$_{5/2}$;
(b) $^4$G$_{5/2}$ $\rightarrow$  $^6$H$_{7/2}$;
(c) $^4$G$_{5/2}$ $\rightarrow$  $^6$H$_{9/2}$;
(d) $^4$G$_{5/2}$ $\rightarrow$  $^6$H$_{11/2}$;
(e) $^4$G$_{5/2}$ $\rightarrow$  $^6$H$_{13/2}$;
(f) $^4$G$_{5/2}$ $\rightarrow$  $^6$F$_{5/2}$;
(g) $^4$G$_{5/2}$ $\rightarrow$  $^6$F$_{7/2}$;
(h) $^4$G$_{5/2}$ $\rightarrow$  $^6$F$_{9/2}$ transitions.
}
\end{figure}

Laser excited fluorescence spectra are shown in Figure \ref{fig:emission}, which were obtained using excitation into the $^{4}$G$_{7/2}$ multiplet at 19957 cm$^{-1}$ with Coumarin 500 dye. Fluorescence to most of the $^{6}$H and $^{6}$F multiplets can be observed with the notable exception of the heavily crystal-field J-mixed $^{6}$H$_{15/2}$, $^{6}$F$_{1/2}$, and $^{6}$F$_{3/2}$ multiplets whose fluorescence appears anomalously weak. Crystal-field levels for these multiplets have been assigned from absorption. From the combination of absorption and laser spectroscopy, a total of fifty-six crystal-field levels have been determined (Table \ref{tab:levels}).

Previous spectroscopic studies of this material \cite{van2012judd,khaidukov2021study} were carried out at room temperature using low-resolution equipment. At room temperature all of the crystal-field levels of the initial multiplet will be populated and the lines will be broadened due to phonon interactions. The numerous broad overlapping lines makes assignment of crystal-field levels impossible. Our low-temperature high-resolution spectra enables identification of the crystal-field levels, which allows a better understanding of the electronic structure.

\clearpage

\subsection{Crystal-Field Analysis}

In our study of Er$^{3+}$ in K$_2$YF$_5$ \cite{Solanki2024},we utilized the magnetic-splitting data from Ref.\cite{Zverev2011}, which enabled us to fit all crystal-field parameters using the computational method outlined in Ref.\cite{horvath2019}. However, due to the absence of magnetic-splitting data for Sm$^{3+}$ doped K$_2$YF$_5$, We adopt the apporach used for Ho$^{3+}$ doped K$_2$YF$_5$  \cite{chanprakhon2025}. To reduce the number of free parameters,  only the crystal-field parameters corresponding to \(q\) = 0 were varied, while those associated with $q \neq 0$ were constrained to follow the ratios previously established for Er$^{3+}$ (Fit B of Table 3 of Ref.\cite{Solanki2024}. In our fit, the ratio values of 1.04 for \textit{B$^2_q$},  0.91 for \textit{B$^4_q$}, and 1.64 for \textit{B$^6_q$} while the ratio values of 1.05 for \textit{B$^2_q$},  0.99 for \textit{B$^4_q$}, and 1.04 for \textit{B$^6_q$} for Ho$^{3+}$ \cite{chanprakhon2025}, The B$^6_q$ are significantly bigger for Sm$^{3+}$ compared to Er$^{3+}$ and Ho$^{3+}$. Starting with the alternative Er$^{3+}$ parameters (Fit A of Table 3 of Ref \cite{Solanki2024}) gave a slightly worse fit, and starting with the Nd$^{3+}$ parameters of Ref.\cite{karbowiak2012energy} gave a much worse fit.

In this study, the free-ion parameters listed in Table \ref{tab:crystalfield} were adjusted, whereas the remaining parameters were taken from Sm$^{3+}$ in LaF$_3$ \cite{Carnall1989}. A total of {56} experimentally assigned energy levels were utilized for the fitting process. The results of the crystal-field fit are presented in Table \ref{tab:levels}, with the corresponding parameters detailed in Table \ref{tab:crystalfield}. The fit had a standard deviation of 16 cm$^{-1}$.
Although this deviation could be reduced by allowing more crystal-field parameters to vary, the absence of magnetic-splitting data makes it impossible to assess the validity of the parameters. Furthermore, since the ratios of the crystal-field parameters were constrained, realistic estimates of uncertainties cannot be provided.

\begin{longtable}{ccr@{\hskip 10mm}r}
\caption{ 
  Experimental and calculated electronic energy levels for Sm$^{3+}$ doped in K$_2$YF$_5$ (in cm$^{-1}$). The experimental uncertainties are 1 cm$^{-1}$ for the data obtained from fluorescence and excitation spectra, and 0.2 cm$^{-1}$ for data obtain from the absorption measurements. The alphanumeric labels follow the standard Dieke convention \cite{Dieke}.}
\label{tab:levels}\\


\hline
Multiplet         &State             & Measured            &Fit \\
\hline

$^6$H$_{5/2}$   &Z$_1$             &             0            &      27.7     \\
                &Z$_2$             &            144           &     142.3     \\
                &Z$_3$             &            283           &     308.9          \\ \\
                   
$^6$H$_{7/2}$   &Y$_1$             &            1097          &     1100.3       \\
                &Y$_2$             &            1162          &     1152.4          \\
                &Y$_3$             &            1272          &     1290.6          \\
                &Y$_4$             &            1381          &     1364.5          \\ \\
                
$^6$H$_{9/2}$   &X$_1$             &            2293.4          &     2307.5          \\
                &X$_2$             &            2373.3          &     2340.7          \\
                &X$_3$             &            2395.7          &     2398.5          \\
                &X$_4$             &            2510.7          &     2520.3          \\
                &X$_5$             &            2580.9          &     2561.2          \\ \\
                
$^6$H$_{11/2}$  &W$_1$             &            3611          &    3615.3          \\
                &W$_2$             &            3668          &    3639.2          \\
                &W$_3$             &            3691          &    3682.1          \\
                &W$_4$             &            3766          &    3768.6          \\
                &W$_5$             &            3845          &    3857.3          \\
                &W$_6$             &            3897          &    3889.2         \\ \\

$^6$H$_{13/2}$  &V$_1$             &            4936          &    4939.2          \\
                &V$_2$             &            5036          &    5012.0          \\
                &V$_3$             &            5098          &    5084.3          \\
                &V$_4$             &            5126          &    5125.1         \\
                &V$_5$             &            5162          &    5161.6          \\
                &V$_6$             &            5263          &    5263.1          \\
                &V$_7$             &            5360          &    5372.4          \\ \\
                
$^6$H$_{15/2}$, &S$_1$             &            6221.0          &    6243.5     \\
$^6$F$_{1/2}$, $^6$F$_{3/2}$            &S$_2$             &            6342.8          &    6336.7          \\
                &S$_3$             &            6409.4          &    6425.0          \\
                &S$_4$             &            6486.5          &    6504.6          \\
                &S$_5$             &             -              &    6600.5          \\
                &S$_6$             &             -              &    6665.5          \\
                &S$_7$             &             -              &    6701.3          \\  
                &S$_8$             &            6740.0          &    6738.7          \\
                &S$_9$             &            6778.1          &    6762.6          \\                          
                &S$_{10}$          &             -            &    6801.6          \\
                &S$_{11}$          &             -            &    6944.1          \\ \\

$^6$F$_{5/2}$   &R$_1$             &            7241.2          &    7219.3          \\
                &R$_2$             &            7278.6          &    7299.2         \\
                &R$_3$             &            7321.7          &    7318.3          \\ \\

$^6$F$_{7/2}$   &Q$_1$             &            8089.3          &     8071.0         \\
                &Q$_2$             &            8116.0          &     8119.7        \\
                &Q$_3$             &            8130.2          &     8131.1         \\
                &Q$_4$             &            8209.2          &     8198.7         \\ \\
                
$^6$F$_{9/2}$   &P$_1$             &            9234.6          &     9242.9         \\
                &P$_2$             &            9276.6          &     9272.7         \\
                &P$_3$             &            9316.0          &     9306.1         \\  
                &P$_4$             &            9325.1          &     9321.9         \\
                &P$_5$             &            9356.3          &     9348.2         \\ \\

$^6$F$_{11/2}$  &O$_1$             &           10567.9          &     10587.8         \\
                &O$_2$             &           10621.6          &     10620.7         \\
                &O$_3$             &           10676.0          &     10672.6         \\
                &O$_4$             &           10695.2          &     10696.1        \\  
                &O$_5$             &           10734.6          &     10738.1         \\
                &O$_6$             &           10773.8          &     10794.7         \\ \\

$^4$G$_{5/2}$   &A$_1$             &           17749            &     17738.4         \\
                &A$_2$             &           -                &     18053.8         \\
                &A$_3$             &           -                &     18162.2         \\ \\

$^4$F$_{3/2}$   &B$_1$             &           18923          &     18932.1        \\
                &B$_2$             &           18954          &     18946.7         \\ \\

$^4$G$_{7/2}$   &C$_1$             &           19957          &     19936.3         \\
                &C$_2$             &           20033          &     20064.2         \\
                &C$_3$             &           20135          &     20160.0         \\
                &C$_4$             &           20282          &     20254.9         \\ \\
\hline
\end{longtable}

\clearpage
\begin{table}[tb!]
\caption{ \label{tab:crystalfield}
  Free-ion and crystal-field parameters for Sm$^{3+}$-doped K$_2$YF$_5$ (in cm$^{-1}$), compared with Ho$^{3+}$ \cite{chanprakhon2025} and Er$^{3+}$ \cite{Solanki2024}.
  Free-ion parameters that were not varied are not shown. Those were fixed to the values obtained for Sm$^{3+}$ in LaF$_3$ \cite{Carnall1989}. 
  Parameters in square brackets were constrained so that the ratios of the parameters with $q \neq 0$ to $q=0$ were the same as for Er$^{3+}$ \cite{Solanki2024}.
}
\footnotesize
\renewcommand{\arraystretch}{1.3}
\begin{tabular}{cc@{\hskip 10mm}c@{\hskip 10mm}c@{\hskip 10mm}c}
\hline
Parameter       &       Sm$^{3+}$     &       Ho$^{3+}$       &       Er$^{3+}$       \\
\hline                                                                
                                                                        
$E_\text{avg}$  &       47608        \\
$F^2$           &       79392        \\   
$F^4$           &       56800        \\   
$F^6$           &       40008        \\    
$\zeta$         &       1170         \\      
$B^2_0$         &       -394          &    -399 &       -379 \\
$B^2_2$         &       [-93-424i]    &       [-94-429i]  &       -89-408i \\ 
$B^4_0$         &       -1153        &       -1250   &       -1267\\ 
$B^4_2$         &       [-66-59i]    &       [-71-64i]  &       -72-65i\\ 
$B^4_4$         &       [-1177-129i]  &       [-1285-141i] &       -1294-142i \\ 
$B^6_0$         &       443        &       281   &       270\\ 
$B^6_2$         &       [220-139i]   &       [140-89i]   &       134-85i\\ 
$B^6_4$         &       [-558-167i]   &       [-354-106i]  &       -340-102i \\ 
$B^6_6$         &       [-315+305i]   &       [-200+194i]  &       -192+186i\\ 

\hline
\end{tabular}
\end{table}
  
\clearpage

\section{Conclusions}
Sm$^{3+}$-doped K$_2$YF$_5$ microparticles, approximately 20 $\mu$m in length, were successfully synthesized with a 2.5\% doping concentration using the hydrothermal technique. A total of {56} experimentally determined crystal-field levels were determined from a combination of infrared absorption and laser excitation and fluorescence, performed at cryogenic temperatures. A parameter constrained crystal-field analysis provides a good approximation to the data.

\section{Acknowledgements}

The authors gratefully acknowledge the technical support provided by Mr.\ Stephen Hemmingson, Mr.\ Graeme MacDonald, Dr.\ Jamin Martin, Mr.\ Matthew Pannell, Dr.\  Matthew Polson, and Mr.\ Robert Thirkettle.
P.C.\ gratefully acknowledges the Royal Thai Government for the Ph.D. scholarship awarded through the Development and Promotion of Science and Technology Talents Project.


\clearpage


\bibliographystyle{elsarticle-num}
\bibliography{SmK2YF5_references.bib}

\end{document}